\documentclass{ifacconf}

\usepackage{graphicx}      % include this line if your document contains figures
\usepackage{natbib}        % required for bibliography
\usepackage{amsmath}       % equation on multiple lines 
\usepackage{amsfonts}      % equation on multiple lines 
\usepackage{float}         % anchor images
\usepackage{verbatim}      % comments 
\begin{document}
\begin{frontmatter}

\title{Hourly operation of a regulated lake via Model Predictive Control} 

%\thanks[footnoteinfo]{Sponsor and financial support acknowledgment
%goes here. Paper titles should be written in uppercase and lowercase
%letters, not all uppercase.}

\author{Raffaele G. Cestari}, 
\author{Andrea Castelletti},
\author{Simone Formentin}

\address{Dipartimento di Elettronica, Informazione e Bioingegneria, Politecnico di Milano, Via G. Ponzio 34/5, 20133, Milan, Italy. Email to: {\tt simone.formentin@polimi.it}, {\tt andrea.castelletti@polimi.it}, {\tt raffaelegiuseppe.cestari@polimi.it}.}

\begin{abstract}                % Abstract of not more than 250 words.
The optimal operation of regulated lakes is a challenging task involving conflicting objectives, ranging from controlling lake levels to avoid floods and low levels to water supply downstream. The traditional approach to operation policy design is based on an offline optimization, where a feedback control rule mapping lake storage into daily release decisions is identified over a set of observational data. In this paper, we propose a receding-horizon policy for a more frequent, online regulation of the lake level, and we discuss its tuning as compared to benchmark approaches. As side contributions, we provide a daily alternative based on the same rationale, and we show that this is still valid under some assumptions on the water inflow. Numerical simulations are used to show the effectiveness of the proposed approach. We demonstrate the  approach on the regulated lake Como, Italy. 
\end{abstract}


\end{frontmatter}
%===============================================================================

\section{Introduction}
In water resources management, advanced control techniques are becoming increasingly popular due to the growing need to adapt to adverse climatic conditions and more impactful extreme weather phenomena. Among the different available approaches, Model Predictive Control (MPC) seems particularly appealing, as it allows to anticipate the evolution of the system and elaborate an optimal control action according to some selected objectives \cite{castelletti-giuliani}. Most of the published contributions consider an MPC scheme based on a linear model, originated from a mass balance equation for reservoirs and lakes or a linearization of the \textit{De Saint-Venant} hydraulic equations for open channels, see, \textit{e.g.}, \cite{Agudelo,raso2014short,Baunsgaard2016,BERKEL2018199,myanmar,turchia,velarde2019scenario}. 
Concerning the formulation of the control objective, most papers propose a quadratic cost (\textit{e.g.}, \cite{Agudelo}, \cite{Baunsgaard2016} and \cite{myanmar}) as it allows to solve the optimization problem efficiently. Choosing a more general nonlinear cost function leads to more flexibility at the cost of higher computational burden, see \cite{raso2014short}, \cite{Galelli}, \cite{ficchi2016optimal}. A full nonlinear MPC is used instead in \cite{Zhou2012}, leading to even better accuracy at the cost of worse computational tractability. 

In the specific case of lake Como considered in this paper,  the operation targets multiple purposes: the lake level must be prevented from  rising above a flooding threshold and dropping below a water shortage threshold while supplying water to four downstream  agricultural districts. The optimal lake operation design is formulated as a \textit{multi-objective optimization problem} (e.g. \cite{castelletti2010}). 

In this paper, we propose the use of an MPC scheme for lake Como. The novel contributions are as follows:
\begin{itemize}
	\item \textit{Novel control scheme:} as far as we are aware, here we propose for the first time to control the level of lake Como using an MPC scheme with hourly historical data and sampling time. 
	\item \textit{Formulation of the dry level avoidance request as a hard constraint}: the asymmetric effects of system disturbance (water inflow) and control action (water release) drove us to the formulation of the Dry Level Avoidance objective as a hard constraint, instead of an objective function. Since the only possible way for the lake level to go below the dry threshold is as a consequence of the control actions, we can downgrade a 3-objective control problem into a 2-objective one.
		\item \textit{Extensive Benchmarking}: we compare the use of linear quadratic implementation and that of a nonlinear cost, showing that the former is advantageous from the computational point of view and guarantees practically equivalent objective performance. We also compare the proposed method with a benchmark strategy, namely Deterministic Dynamic Programming, providing the maximum offline performance. A sensitivity analysis of the main hyperparameters of the algorithm is also performed.
	\item \textit{Daily approximation of hourly MPC}: In practice, the availability of hourly water inflow data is not guaranteed. This depends on the difficulty of estimating via water mass balance inversion the net inflow to the lake, e.g. due to the presence of seiches. For this reason, we developed a \textit{daily} approximator of the above (nominal) {hourly} MPC. Under the assumption of gaussian intra-day water inflow dynamics, such a daily approximation is proven to have comparable performance. 
\end{itemize}

The remainder of the paper is as follows. In Section \ref{sec:problem}, the lake Como system is described and the control problem is formally stated.
Section \ref{mpc} describes the proposed MPC strategy, while Section \ref{sec:results} illustrates some experimental results obtained on a realistic simulator of the lake dynamics. The paper is ended with some concluding remarks.

\section{System description and problem statement}\label{sec:problem}
\subsection{The Lake Como system}
\label{Como}
\begin{figure}
	\begin{center}
		\includegraphics[width=8.4cm]{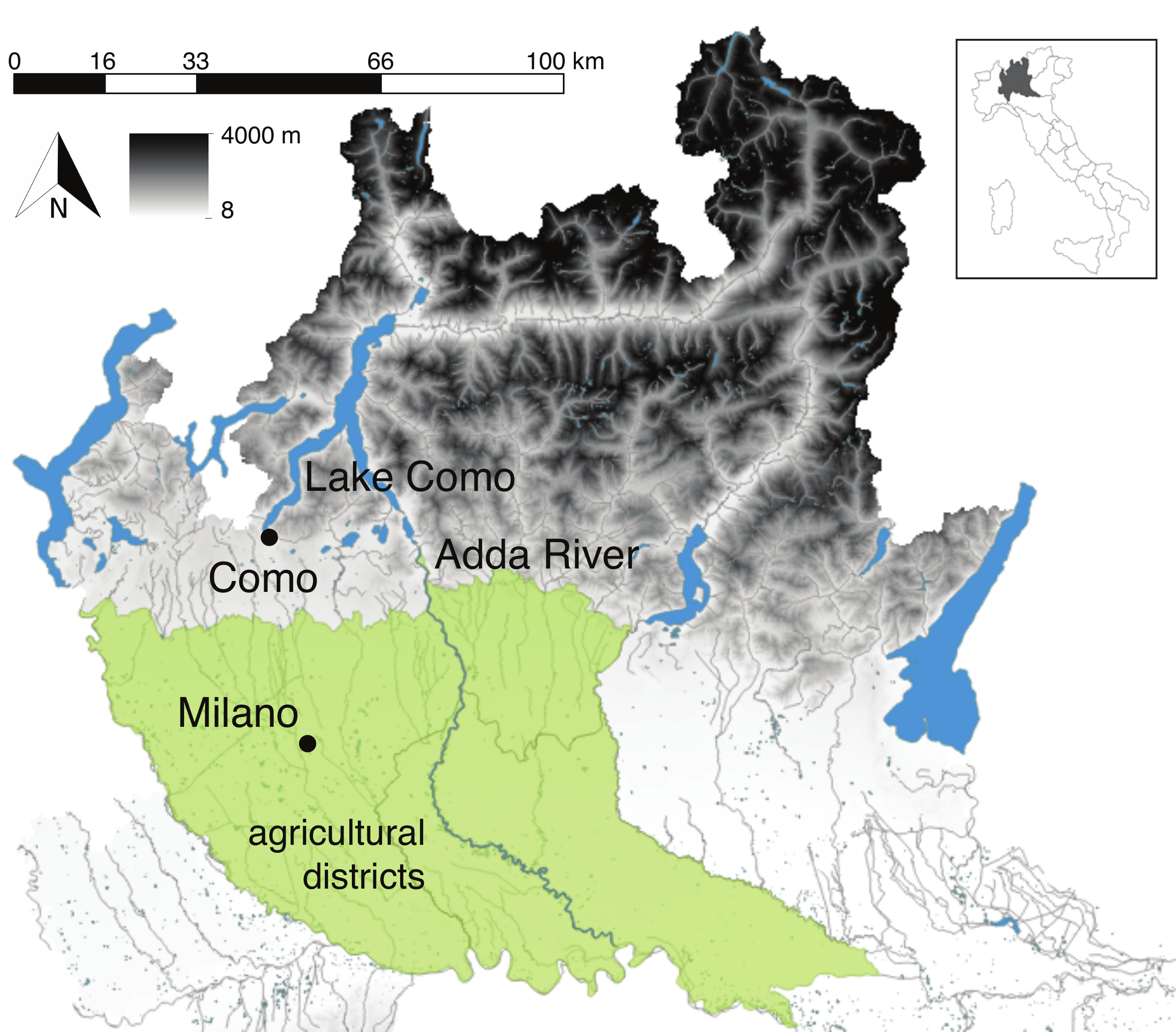}    % The printed column width is 8.4 cm.
		\caption{Lake Como Basin in Northern Italy} 
		\label{fig:Como}
	\end{center}
\end{figure}

Lake Como is located in northern Italy, Lombardy, a region characterized by the abundance of water, lakes, and rivers, see Figure \ref{fig:Como}. As described in \cite{giuliani2019detecting}, lake Como is a subalpine lake, the third-largest lake in Italy and, reaching a maximum depth of $410 \,\, m$, the fifth-deepest lake in Europe. The Adda river is the greatest tributary and also serves  as the lake outlet. The lake catchment covers $4733 \,\, km^2$. Referring to \cite{denaro2017informing} its total volume is $23.4 \,\, km^3$ of which $254 \,\, Mm^3$ can be controlled through the Olginate dam. The lake serves a dense network of irrigation canals, which ultimately feed four agricultural districts covering an area of $1400\,\, km^2$, mainly cultivated with corn. 

One of the major problems in the operation of lake Como, as for many other reservoirs and lakes, is the uncertainty related to the water inflow. Indeed, while water release can be measured with high frequency and accuracy, the inflow cannot be directly measured, as there are 37 tributaries, distributed all along the lake shores. Moreover, the inflow is driven by several exogenous factors, such as rainfall, snow-melt and water releases from upstream hydropower dams. In particular, the dams releases are independent of the lake Como control decisions, being guided mainly by energy market fluctuations. For all these reasons, the water inflow can only be computed \textit{a-posteriori}, by inverting the mass balance equation of the lake Como reservoir. This calls for a responsive operation management, which is able to adapt the control action in real-time depending on the mismatch between the measurements and the expectations.

%%%%%%%%%%%%%%%%%%%%%%%%%%%%%%%%%%%%%%%%%%%%%%%%%%%%%%%%%%%%%%%%%%%%%%%%%%%%%%%%%%%%%%%%%%%%%%%%
\subsection{Control Objectives}
\label{obiettivi}
The control objective considered in this paper is to regulate the level of lake Como by simultaneously satisfying the downstream irrigation demand, avoiding lake floods and too low lake levels. This control problem can be formally stated by defining the following 3 objectives: 
\begin{enumerate}
	\item \textit{Flood avoidance}: the city of Como, located on the south-west side of the lake, is subject to flood events striking the main city square, leading to damages and economic consequences. The flood occurrences, i.e. lake levels higher than $h_F=1.1 \,\, m$, must then be minimized.  
	\item \textit{Dry avoidance}: at the same time, the lake level should also not go below a relative level of $h_D=-0.2 \,\, m$. This objective is of paramount importance for  navigation as too low levels can impact piers and docks and impede the boarding of people and goods, as well as for aesthetic reasons. Since lake Como is a popular touristic venue, dry conditions might lead to an economic rebound in terms of attractiveness of the area. 
	\item \textit{Satisfaction of water demand}: as stated in Section \ref{Como}, the agricultural district demands a specific water flow rate varying throughout the year. A deficit in demand satisfaction may lead to significant economic losses in terms of lack of agricultural production.
\end{enumerate}
In the literature, see e.g. \cite{denaro2017informing}, the objectives usually considered for lake Como water management are flood avoidance and water demand satisfaction. The objective of dry avoidance has been recently introduced to account for climate change, which induced increase in droughts and water scarcity. 

The complexity of the overall problem arises from the difficulty in balancing the above 3 goals.
During spring and summer seasons, snowmelt generates water inflow useful for agriculture, however, in the same period spring rainfalls might cause lake level rising. It follows that the excess water should be stored, but at the same time, it needs to be released to avoid floods. Also, dry avoidance competes with water demand satisfaction as, in summer, when the highest water flow rate is required, the lake level is likely to go below the dry level, leading to a harmful water shortage condition. In the next section, we will discuss how to formally state and solve this problem according to a receding-horizon rationale.

\section{MPC for Lake Como}\label{mpc}
\subsection{Mathematical modeling}
\label{system}
The dynamics of the lake Como is described by a mass balance phenomenon. In continuous time, this can be expressed as
\begin{equation}
	\frac{ds(t)}{dt} = q(t)-r(t),
	\label{continuousMB}
\end{equation}
where $s \,\, [m^3]$ represents the lake Como \textit{storage}, $q \,\, [m^3/s]$ is the \textit{water inflow} and $r \,\, [m^3/s]$ the \textit{water release}.
The evaporation term is already included in the net lake inflow, because the regulated lake portion is almost cylindrical, as for most of the sub-alpine lakes. In the regulation of other water basins, this might not be done as evaporation depends on the lake surface.

In discrete time, Equation \eqref{continuousMB} reads as
\begin{equation}
	s(t_h) = s(t_h-1) + 3600(q(t_h)-r(t_h)),
\end{equation}
where $t_h$ denotes a \textit{hourly} time index.
The lake level $h$ turns out to be given by the algebraic relationship
\begin{equation}
	h(t_h)=\frac{s(t_h)}{A} + h_0,
\end{equation}
where $A = 145900000 \,\,m^2$ is the lake Como surface and $h_0=-0.4 \,\, m$ is a fixed reference value to compute the relative lake level. Notice that the water inflow $q(t_h)$ here represents a \textit{disturbance} that \textit{cannot} be directly measured because of the reasons mentioned in Section \ref{Como}. The water release $r(t_h)$ is also saturated as follows:
\begin{equation}	
	r(t_h) =
	\left \{ \begin{array}{rl}
		&r_m(t_h) \,\,\, if \,\, u(t_h) \leq r_m(t_h)\\
		&r_M(t_h) \,\,\, if \,\, u(t_h) \geq r_M(t_h)\\
		&u(t_h) \,\,\, if \,\, r_m(t_h)<u(t_h)<r_M(t_h)
	\end{array}
	\right.
	\label{release}
\end{equation}
where $u(t_h)$ is the water release computed by the controller, $r_m(t_h)$ and $r_M(t_h)$ are the lower and upper physical bounds on lake release. 
The bounds follow the relationships
\begin{equation}
	r_m(t_h) = 
\left \{ \begin{array}{rl}
	&0 \,\,\, if \,\, h(t_h) \leq h_0\\
	&MEF \,\,\, if \,\, h(t_h) \leq h_F\\
	&k(h(t_h) + n)^{e} \,\,\, if \,\, h(t_h)>h_F,
\end{array}
\right.
\label{minRelease}
\end{equation}
\begin{equation}
	r_M(h) = 	
	\left \{ \begin{array}{rl}
		&0 \,\,\, if \,\, h(t_h) \leq h_0\\
		&k(h(t_h) + n)^{e} \,\,\, if \,\, h(t_h)>h_0,
	\end{array}
	\right.
	\label{maxRelease}
\end{equation}
where $k = 33.37,\, n = 2.5,\,e = 2.015$ are \textit{a-priori} given parameters determined by empirical characterization of the actuation system.
The lower bound $r_m(t_h)$ must guarantee the \textit{Minimum Environmental Flow} $MEF = 10\,\, m^3/s$ to preserve the river ecosystem. 
The upper bound $r_M(t_h)$ must satisfy physical constraints due to the shape of the dam. 
Equations \eqref{release}, \eqref{minRelease} and \eqref{maxRelease}  describe the nonlinear behaviour affecting water release. From Equation \eqref{release}, it is clear that the control action must comply with the system constraints, which makes MPC a good candidate strategy for on-line optimization, for its possibility of including hard signal bounds. 

\begin{rem}
	Traditionally, the case study of lake Como has been approached with a daily sampling period. To resort to that system representation, the daily mass balance (where $t_d$ is day time index) could be derived from an integration of the hourly dynamics,
\begin{equation}
	s(t_d+1)=s(t_d)+ 3600\left(\sum_{t_h=0}^{23} {q(t_h)} - \sum_{t_h=0}^{23} {r(t_h)}\right)
	\label{sDay}
\end{equation}
whereas the daily water release could be computed as the average water release over the 24 hours. 
\end{rem}

\subsection{Receding-horizon control}
\label{MPCstruct}
\begin{comment}
\begin{figure} [H]
	\begin{center}
		\includegraphics[width=8.3cm]{HQHourMPC}    % The printed column width is 8.4 cm.
		\caption{Hourly MPC Controller Scheme} 
		\label{fig:HMPC}
	\end{center}
\end{figure}
\end{comment}
In this paper, we formulate the MPC problem starting from the linear quadratic formulation already proposed in \cite{turchia}, \cite{Galelli}, and \cite{ficchi2016optimal}. 
%Its application in water resources is well-supported in literature, see \cite{myanmar}, \cite{Baunsgaard2016} and \cite{Agudelo}. An example of this formulation, described in \cite{myanmar}, is 
%\begin{subequations}
%	\begin{equation}
%		\operatorname{min}_{u}  \,\,\sum_{t_h=0}^{H}{Z(t_h)^TWZ(t_h) + U(t_h)^TRU(t_h)}
%	\end{equation}
%	\begin{equation}
%		Z(t_h+1) = AZ(t_h)+B_uU(t_h) + B_dD(t_h) 
%	\end{equation}
%	\begin{equation}
%		U_{Min} \leq U(t_h) \leq U_{Max}\end{equation}
%	\begin{equation}
%		Z_{Min} \leq Z(t_h) \leq Z_{Max}     
%	\end{equation}
%	\label{linearLoro}
%\end{subequations}
%where $Z(t_h)$ is a vector of states of the system (lake level, conduit outflow, lake level deviation from reference), $U(t_h)$ is a vector of control actions (variation of conduit outflow between consecutive time steps), $D(t_h)$ is a vector of disturbances (water inflow), $A,B_u,B_d$ are system matrices, $W$ and $R$ are the weighted matrices defining the relative importance of sub-objectives within the optimization process. $Z_{Min}$, $Z_{Max}$, $U_{Min}$, $U_{Max}$ are the lower (Min) and upper (Max) values that the states $Z$ and the control action $U$ can assume. The same structure can be also found in \cite{velarde2019scenario}. 

Specifically, the model predictive controller employed in this work has the following form:
\begin{comment}
\begin{equation}
	\begin{split}
		&min_u \,\,\sum_{t_h=0}^{H-1} {||\epsilon_{Max}||_2+\lambda||\epsilon_{Demand}||_2}  \\
		&subject \,\, to \\
		&s(t_h+1) = s(t_h) +3600(q(t_h)-u(t_h)) \\
		&u_{Min} \leq u(t_h) \leq u_{Max} \\
		&u(t^*) \geq w(t_h) +\epsilon_{Demand} \\
		&s_{Min} \leq s(t_h) \leq s_{Max} + \epsilon_{Max}\\ 
		&\epsilon_{Max} \geq 0,
	\end{split}
\end{equation}
\end{comment}
\begin{subequations}
		\label{MPCmio}
	\begin{equation}
		\operatorname{min}_{u} 
		\,\,\sum_{t_h=0}^{H-1} {||\epsilon_{Max}||_2+\lambda||\epsilon_{Demand}||_2}
	\end{equation}    
	\begin{equation}
		s(t_h+1) = s(t_h) +3600(q(t_h)-u(t_h)) 
	\end{equation}
	\begin{equation}
		u_{Min} \leq u(t_h) \leq u_{Max}
	\end{equation}
	\begin{equation}
		u(t_h) \geq w(t_h) +\epsilon_{Demand} \\
	\end{equation}
    \begin{equation}
		s_{Min} \leq s(t_h) \leq s_{Max} + \epsilon_{Max}
		\label{hardCnst}
	\end{equation}
	\begin{equation} 
			\epsilon_{Max} \geq 0
	\end{equation}
	\begin{equation}
			H = 24
	\end{equation}
\end{subequations}

where $w(t_h)$ is the agricultural water demand,  $H$ is the prediction horizon of $24$ hours. This choice is reasonable since it allows to take into account both hourly and daily dynamics in the optimization. $s_{Min}$, $s_{Max}$, $u_{Min}$, $u_{Max}$ represent the lower (Min) and upper (Max) values that the lake storage $s$ and the control action $u$ can assume. Control action boundaries are updated at each time step and coincide with the ones specified in \eqref{minRelease} and \eqref{maxRelease}, where $u_{Max}=r_m(t_h)$ and $u_{Min}=r_m(t_h)$. 

Notice that the approach in \eqref{MPCmio} is \textit{deterministic} as the disturbance, namely the water inflow $q$, is assumed to coincide with the actual realization of the disturbance over the prediction horizon, thus generating a unique trajectory for the state of the system.

The problem formulation \eqref{MPCmio} considers the goal of flood avoidance through $\epsilon_{Max}$ and that of water demand satisfaction through $\epsilon_{Demand}$. These objectives are defined as \textit{soft constraints} as a hard formulation might lead to infeasible solutions for certain values of the disturbance.  
Alternative formulations available in literature, like the ones in \cite{Galelli} and \cite{ficchi2016optimal}, differ from the one presented in this work because they formulate the control objective using nonlinear cost terms, \textit{e.g.} $max((h_l - h_F),0)^2$, $max((w - r),0)^2$. Such formulations are definitely more accurate, but ask for a nonlinear solver, which requires higher computational time. In this work, we will show that the difference in performance does not justify the use of more complex cost functions.

One novel contribution of the proposed MPC structure \ref{MPCmio} is imposing the dry avoidance objective as a \textit{hard constraint}, see again Equation \eqref{hardCnst}. This formulation is driven by the physical interpretation of the lake Como system. Going below the dry level is possible only as a consequence of a specific control action. This means that the dry avoidance objective can always be fulfilled without explicitly adding a related cost term. In this way, only 2 objectives need to be traded off in the cost. This simplifies the design when tuning the parameter $\lambda$, as it will be clear in Section \ref{sensitivity}. 
%%%%%%%%%%%%%%%%%%%%%%%%%%%%%%%%%%%%%%%%%%%%%%%%%%%%%%%%%%%%%%%%%%%%%%%%%%%%%%%%%%%%%%%
\section{Simulation results}\label{sec:results}
The MPC strategy presented in this paper is  assessed within a simulation environment. Specifically, the simulator is built using the mathematical model of Section \ref{system}, while the controller employs a linear system description, as illustrated in \eqref{MPCmio}. To work on a fairly realistic scenario, we employ the real data of the daily inflow of the year $2000$ and simulate over a horizon of $365$ days ($8760$ hours). All computations are carried out on an Intel Core i7-8750H with 6 cores, at 2.20 GHz (maximum single core frequency: 4.10 GHz), 16 GB RAM, running Matlab R2019b. To start with, we will provide a comparative analysis with respect to the benchmark DDP. Then, we will show that the daily version of our predictive strategy can be used as a good proxy in terms of overall performance. Finally, we will discuss pros and cons of the performance of the proposed MPC as compared to a nonlinear formulation.

\subsection{MPC vs DDP}
\label{sensitivity}
Deterministic Dynamic Programming\footnote{For the interested reader, see \cite{perera1996reservoir} and \cite{faber2001reservoir} for a stochastic dynamic programming version.} is an offline method that computes the optimal value function and derives the optimal control policy by performing backwards dynamic optimization. This offline approach defines an ideal optimal control because it considers all the available data in an anti-causal fashion. Moreover, the disturbance realization is the actual one, namely the inflow is supposed to be perfectly known. 

In particular, we will consider a DDP setting as follows: 
\begin{enumerate}
	\item The objective weights are selected respectively as $w_{Flood} = 0.4$, $w_{Demand} = 0.6$, $w_{Dry} = 0$. They are chosen according to a specific compromise among the 3 objectives, selecting a corner point of the Pareto Front as indicated in \cite{denaro2017informing}.  
	\item The policy is optimized over the period 1/1/1999 - 31/12/2019.
\end{enumerate}

\begin{figure} [H]
 	\begin{center}
 		\includegraphics[width=8.4cm]{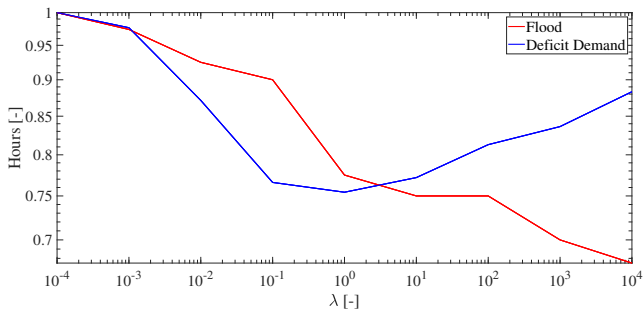}    % The printed column width is 8.4 cm.
 		\caption{Sensitivity Analysis with respect to $\lambda$.} 
 		\label{fig:lambda}
 	\end{center}
 \end{figure}

 DDP is compared to the proposed MPC method, implemented as in \eqref{MPCmio}. For a proper tuning of $\lambda$, a sensitivity analysis  has been performed and illustrated in Figure \ref{fig:lambda}. In particular, in the figure, the (normalized) number of flood hours is contrasted to the (normalized) number of deficit demand hours for several configurations of $\lambda$, ranging from $10^{-4}$ to $10^4$. The two considered metrics, according to which the optimal $\lambda$ has been chosen, are representative of the control objectives of flood avoidance and agricultural water demand satisfaction. While the flood hours decrease as $\lambda$ increases, demand hours have their minimum at $\lambda=1$. From these observations, the selection of $\lambda=1$ follows. 

\begin{figure*}[h!]
	\begin{center}
		\includegraphics[scale=0.35]{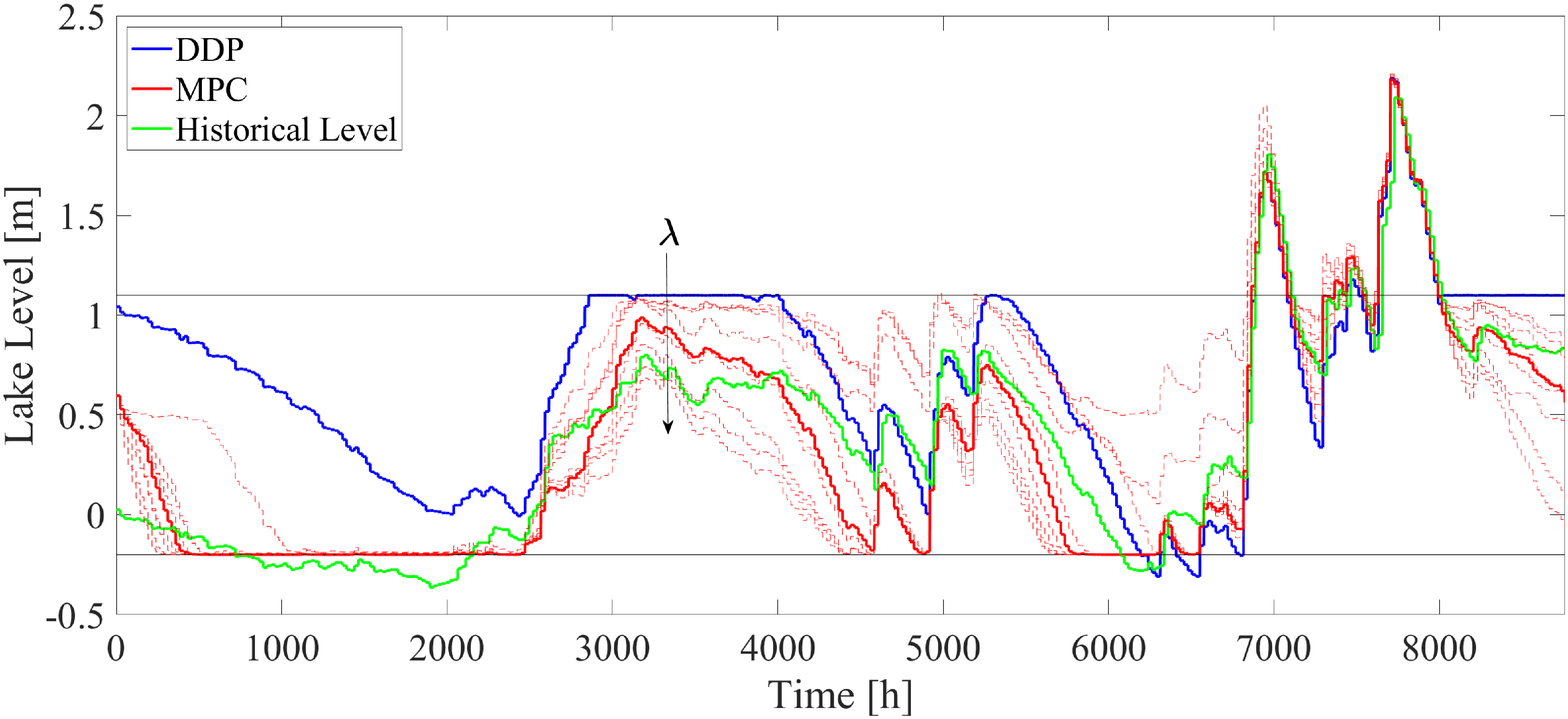}    % The printed column width is 8.4 cm.
		\caption{Time trajectory of the Lake Como level according to different strategies: DDP (blue), Historical Level (green) and MPC with varying $\lambda$ (red). The best choice $\lambda=1$ is denoted by the solid line, while the others are indicated by the dashed lines.} 
		\label{fig:hAll}
	\end{center}
\end{figure*}

 Figure \ref{fig:hAll} shows the time evolution of the lake level. Specifically, the red curve represents the MPC with $\lambda =1$, while all the other red-dashed curves show the variation of the level for varying $\lambda$. The red-dashed curves tend to flatten against the lower bound as $\lambda$ increases, being even more reactive against flood occurrence. However, all of them do not violate the hard constraint on the lower lake level. The blue curve represents the lake level obtained with offline DDP. The green curve is the historical lake level. In the initial part of the year, from $0$ to $2500\,\, [h]$, the lake level of DDP is significantly higher. This derives from an offline knowledge of the whole simulation horizon that allows one to better satisfy the irrigation demand by graduating the water release, also because of the knowledge of the inflow trends. This more economical management of the releases allows to satisfy the irrigation demand, as also indicated in Table \ref{tb:MPCvsDDP}. In the same period, however, the reader can notice how the red curve is able to fulfill the dry level constraint perfectly, while the historical regulation fails in preventing the lake from going into a dry condition. This fact proves the effectiveness of the proposed strategy. Later, the blue and nominal red curves tend to overlap for $ \lambda = 10^{-4},10^{-3}$. Finally, after the $7700th$ hour, the curves are practically overlapping. This depends on the high level condition characterizing this phase. 
 Concerning the objectives, the best MPC configuration is comparable with DDP considering the flood objective. Looking at demand satisfaction, the offline approach is clearly working better, at the cost of worse performances in terms of the dry objective. 

 \begin{comment}
 \begin{figure} [H]
 	\begin{center}
 		\includegraphics[width=8.4cm]{hAllNewConStorico}    % The printed column width is 8.4 cm.
		\caption{Time trajectory of the Lake Como level according to different strategies: DDP (blue), Historical Level (green) and MPC with varying $\lambda$ (red). The best choice $\lambda=1$ is denoted by the solid line, while the others are indicated by the dashed lines.} 
		\label{fig:hAll}
 	\end{center}
 \end{figure}
\end{comment}

\begin{comment}
\begin{figure} [H]
	\begin{center}
		\includegraphics[width=8.3cm]{rAll}    % The printed column width is 8.4 cm.
		\caption{Hourly MPC Water Release varying $\lambda$ vs DDP $[m^3/s]$} 
		\label{fig:rAll}
	\end{center}
\end{figure}
\end{comment}
\begin{table}[H]
	\begin{center}
		\caption{MPC vs DDP}\label{tb:MPCvsDDP}
		\begin{tabular}{|p{3.5cm}|p{2cm}|p{2cm}|}
			\hline
			&\textit{MPC} &\textit{DDP} \\
			\hline
			%\textit{Flood Objective}& & \\ \hline
			\textit{RMSE Flood} $[m]$ & 0.5215 & 0.5529 \\
			\hline
			\textit{Lake Level Peak} $[m]$& 2.178 & 2.187\\ 
			\hline
			\textit{Flood Hours} & 744 & 624 \\
			\hline
			\textit{Area Flood} $[m*hours]$ & 316.160 & 289.740 \\
			\hline 
			%\textit{Demand Objective}& & \\ \hline
			\textit{RMSE Demand} $[m^3/s]$ & 53.200 & 15.083 \\
			\hline
			\textit{Deficit Peak} $[m^3/s]$& -144.190 &-25.560\\ 
			\hline
			\textit{Deficit Hours} & 3096 & 4584 \\
			\hline
			\textit{Area Deficit} $[m^3/s*hours]$ & 131570 & 61667 \\
			\hline 
			%\textit{Dry Objective}& & \\ \hline
			\textit{RMSE Dry} $[m]$ & 0 & 0.070 \\
			\hline
			\textit{Lake Level Minimum} $[m]$& -0.2 & -0.310\\ 
			\hline
			\textit{Dry Hours} & 0 & 624 \\
			\hline
			\textit{Area Dry} $[m*hours]$ & 0 & 15.910 \\
			\hline
		\end{tabular}
	\end{center}
\end{table}
\begin{comment}
\begin{table}[H]
	\begin{center}
		\caption{Hourly MPC vs DDP Demand Performances}\label{tb:DemandDDP}
		\begin{tabular}{|p{3.5cm}|p{2cm}|p{2cm}|}
			\hline
			&\textit{MPC} &\textit{DDP} \\ \hline
			\textit{RMSE Demand} $[m^3/s]$ & 53.200 & 15.083 \\
			\hline
			\textit{Deficit Peak} $[m^3/s]$& -144.190 &-25.560\\ 
			\hline
			\textit{Deficit Hours} & 3096 & 4584 \\
			\hline
			\textit{Area Deficit} $[m^3/s*hours]$ & 131570 & 61667 \\
			\hline 
		\end{tabular}
	\end{center}
\end{table}
\begin{table}[H]
	\begin{center}
		\caption{Hourly MPC vs DDP Dry Performances}\label{tb:DryDDP}
		\begin{tabular}{|p{3.5cm}|p{2cm}|p{2cm}|}
			\hline
			&\textit{MPC} &\textit{DDP} \\ \hline
			\textit{RMSE Dry} $[m]$ & 0 & 0.070 \\
			\hline
			\textit{Lake Level Minimum} $[m]$& -0.2 & -0.310\\ 
			\hline
			\textit{Dry Hours} & 0 & 624 \\
			\hline
			\textit{Area Dry} $[m*hours]$ & 0 & 15.910 \\
			\hline 
		\end{tabular}
	\end{center}
\end{table}
\end{comment}

\subsection{Hourly MPC vs Daily MPC}
\label{oragiorno}
In this section, we compare the performance of the considered hourly sampling setting with a more traditional setting where new data are available once a day. Both the MPC architectures solve the same optimal control problem in \eqref{MPCmio} with $\lambda=1$. In the hourly case, MPC works according to the receding-horizon principle and $H=24$. The first computed control action is applied to the system, while the others are thrown away, as the optimization is run the following hour again. In the daily case, the computed control actions are applied \textit{in open-loop} over the following 24 hours, as no new data are available.

\begin{figure}[h!]
	\begin{center}
		\includegraphics[width=7cm]{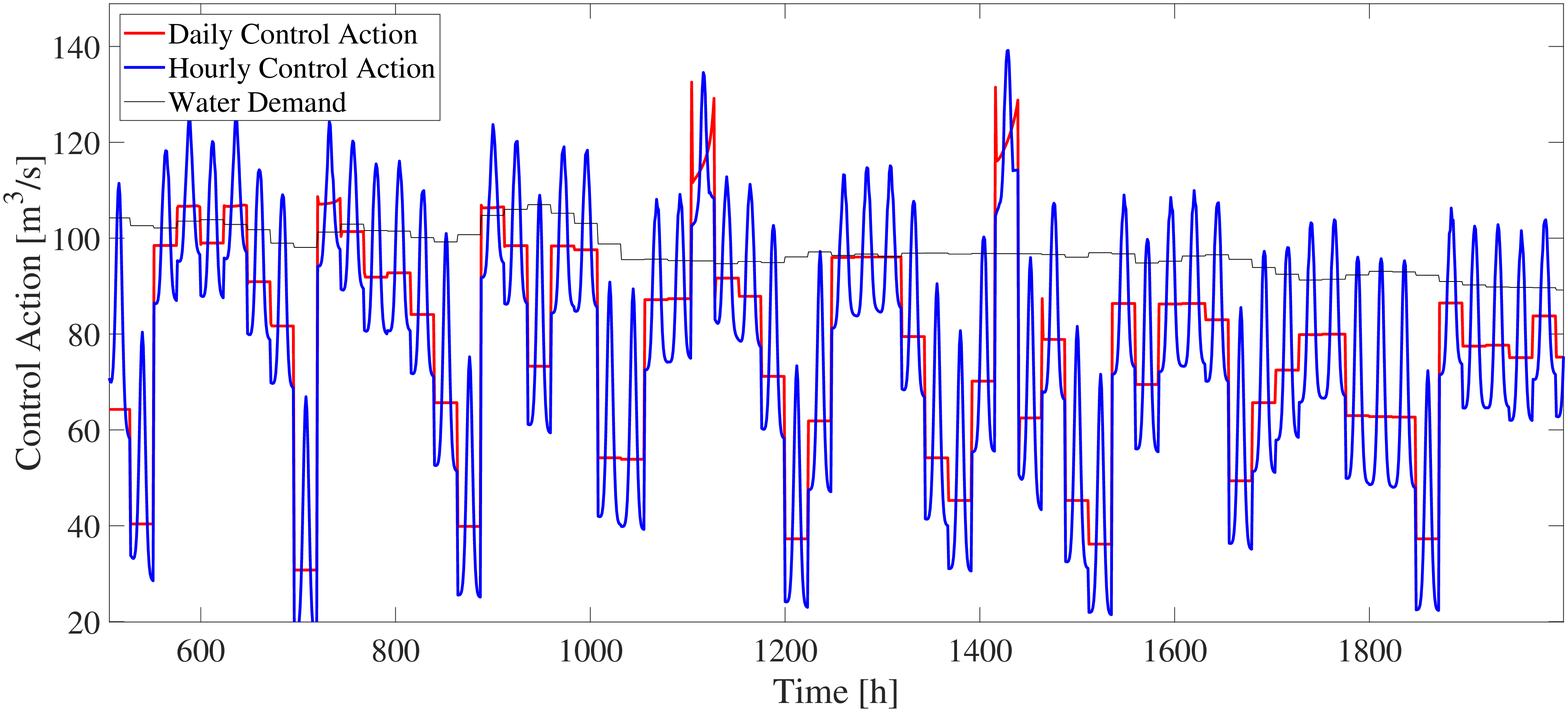}    % The printed column width is 8.4 cm.
		\end{center}
		\caption{Time trajectories of the control actions for the daily and the hourly versions of the proposed MPC method.} 
		\label{fig:uDayVsHourZoomed}
\end{figure}

To highlight the potential differences between the two implementations, an artificial hourly water inflow trajectory is created. This is generated as the sum of the real daily measured water inflow extended constantly over 24 hours $q_{day}(t_h)$, plus an additional artificial Gaussian term $q_{add}(t_h)$ to simulate a plausible bell-shaped \textit{intra-day} inflow dynamics. The water inflow expression is then described as
\begin{equation}
	\begin{split}
		&q(t_h) = q_{day}(t_h) + q_{add}(t_h)\\
		&q_{add}(t_h)= ae^{-b(t_h-h^{Mid})},
	\end{split}
	\label{inflow}
\end{equation}
where $a = 50$, $b=0.06$, $h^{Mid} = 12$ are parameters used to generate the bell-shaped intra-day inflow dynamics. The differences between the two strategies are evident in Figure \ref{fig:uDayVsHourZoomed}, where the computed control actions are compared.
\begin{comment}
\begin{figure*}[h!]
	\begin{center}
		\includegraphics[scale=0.3]{uDayVsHourZoomed}    % The printed column width is 8.4 cm.
		\end{center}
		\caption{Time trajectories of the control actions for the daily and the hourly versions of the proposed MPC method.} 
		\label{fig:uDayVsHourZoomed}
\end{figure*}
\end{comment}
More specifically, the {hourly MPC} control action is more reactive and can better track the  inflow following the intra-day dynamics. The {daily MPC} control action instead captures only the average water inflow behaviour. Nonetheless, the performances between the daily and the hourly settings are comparable both in terms of flood and demand objectives. In both cases, the hard constraint formulation of the dry objective is respected. Therefore, the {daily MPC} implementation could be considered as a {daily} approximation of the {hourly MPC} implementation. Indeed, it is not always guaranteed to have available data with hourly granularity. Hence, one might think of using the {daily} implementation instead of the {hourly} one, under the assumption of an inflow dynamics as the one described in Equation \eqref{inflow}. The approximating {daily MPC} works because the disturbance intra-day dynamics effect on the system (which is not available to the {daily} implementation) is embedded within the {lake storage} information, which is updated every 24 hours. In this way, the {daily MPC} learns the impact of the water inflow on the system in the previous 24 hours. 

Still, a small decrease in performance is noticed. This depends on both a slower response of the controller and on a time-varying saturation-dependent phenomenon. In particular, the MPC receives upper and lower bounds for control action hourly according to Equations \eqref{minRelease} and \eqref{maxRelease} only in the case of the hourly MPC, while every 24 hours otherwise. This leads to the fact that MPC may compute a control action lower than a physically feasible one, because the real saturation is higher than the computed one and viceversa. The combination of the two mentioned problems leads to an unavoidable decrease in performance as shown in Table \ref{tb:HourVsDay}. 

\begin{comment}
\begin{figure} [H]
\begin{center}
\includegraphics[width=8.3cm]{hDayVsHour}    % The printed column width is 8.4 cm.
\caption{Daily vs Hourly MPC Lake Level $[m]$} 
\label{fig:hDayVsHour}
\end{center}
\end{figure}
\begin{figure} [H]
\begin{center}
\includegraphics[width=8.3cm]{rDayVsHour}    % The printed column width is 8.4 cm.
\caption{Daily vs Hourly MPC Water Release $[m^3/s]$} 
\label{fig:rDayVsHour}
\end{center}
\end{figure}

\begin{figure*}[h!]
	\begin{center}
		\includegraphics[scale=0.4]{LevelAndReleaseHourAndDay}    % The printed column width is 8.4 cm.
		\caption{Time trajectories of the lake level and water release for the daily and the hourly versions of the proposed MPC method. The values of the thresholds and water demand are also illustrated to highlight the trade-off among the different costs.} 
		\label{fig:LevelAndReleaseHourAndDay}
	\end{center}
\end{figure*}
\end{comment}

\begin{figure}[h!]
	\begin{center}
		\includegraphics[width=8.4cm]{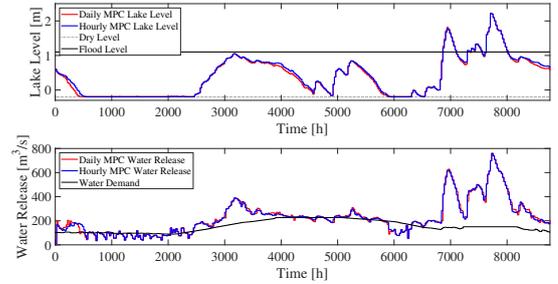}    % The printed column width is 8.4 cm.
		\caption{Time trajectories of the lake level and water release for the daily and the hourly versions of the proposed MPC method. The values of the thresholds and water demand are also illustrated to highlight the trade-off among the different costs.} 
		\label{fig:LevelAndReleaseHourAndDay}
	\end{center}
\end{figure}

\begin{table}[h!]
	\begin{center}
		\caption{Daily vs Hourly MPC.}\label{tb:HourVsDay}
		\begin{tabular}{|p{3.5cm}|p{2cm}|p{2cm}|}
			\hline
			&\textit{Daily} &\textit{Hourly} \\ \hline
			%\textit{Flood Objective}& & \\ \hline
			\textit{RMSE Flood} $[m]$ & 0.527 & 0.513 \\
			\hline
			\textit{Lake Level Peak} $[m]$& 2.221 & 2.215\\ 
			\hline
			\textit{Flood Hours} & 864 & 888 \\
			\hline
			\textit{Area Flood} $[m*hours]$ & 360.800 & 357.810 \\
			\hline 
			%\textit{Demand Objective}& & \\ \hline
			\textit{RMSE Demand} $[m^3/s]$ & 40.876 & 40.342 \\
			\hline
			\textit{Deficit Peak} $[m^3/s]$& -118.390 &-116.480\\ 
			\hline
			\textit{Deficit Hours} & 2688 & 2448 \\
			\hline
			\textit{Area Deficit} $[m^3/s*hours]$ & 79244 & 71688 \\
			\hline 
		\end{tabular}
	\end{center}
\end{table}

\subsection{Linear quadratic vs nonlinear costs}
\label{tempo}
The objective performances achieved using MPC with linear quadratic (LQ) cost and nonlinear costs are almost equivalent and they are not shown for the sake of space. Instead, the requested computational times are compared in Table \ref{lq_vs_nl}. For the LQ case, CVX SDPT3 solver is used, whereas the nonlinear case is implemented using the NLMPC Matlab-Simulink block, by default using the \textit{fmincon} function.

\begin{table}[H]\label{lq_vs_nl}
	\begin{center}
		\caption{MPC with LQ cost vs NL cost: average optimization time over $8760$ iterations.}\label{tb:cputime}
		\begin{tabular}{|p{3.5cm}|p{2cm}|p{2cm}|}
			\hline
			&\textit{LQ} &\textit{NL} \\ \hline
			\textit{Average Time} $[s/call]$ & 0.344 & 1.091 \\
			\hline
		\end{tabular}
	\end{center}
\end{table}
\begin{comment}
\begin{table}[H]
\begin{center}
\caption{LQ MPC vs NLC MPC Objective Performances}\label{tb:performancesNL_L}
\begin{tabular}{|p{3.5cm}|p{2cm}|p{2cm}|}
\hline
&\textit{LQ MPC} &\textit{NLC MPC} \\ \hline
%\textit{Flood Objective}& & \\ \hline
\textit{RMSE Flood} $[m]$ & 0.523 & 0.518 \\
\hline
\textit{Lake Level Peak} $[m]$& 2.110 & 2.140\\ 
\hline
\textit{Flood Hours} & 624 & 741 \\
\hline
\textit{Area Flood} $[m*hours]$ & 273.685 & 314.432 \\
\hline 
%\textit{Demand Objective}& & \\ \hline
\textit{RMSE Demand} $[m^3/s]$ & 57.406 & 58.961 \\
\hline
\textit{Deficit Peak} $[m^3/s]$& -180.240 & -137.704\\ 
\hline
\textit{Deficit Hours} & 3769 & 3938 \\
\hline
\textit{Area Deficit} $[m^3/s*hours]$ & 173550 & 193980 \\
\hline 
\end{tabular}
\end{center}
\end{table}
\end{comment}
Clearly, LQ MPC guarantees a faster solution than the nonlinear implementation.
Some small differences in time evaluation can be actually measured due to the different sources of time data. In fact, in the LQ MPC case, the computational time is taken as one output of the optimization, whereas in the NLC MPC case as the output time of the Simulink block. However, such differences are not relevant since the time required for the simulation is negligible with respect to that required for optimization.
%%%%%%%%%%%%%%%%%%%%%%%%%%%%%%%%%%%%%%%%%%%%%%%%%%%%%%%%%%%%%%%%%%%%%%%%%%%%%%%%%%%%%%%%%%%%%%%%%
\section{Conclusions and Future Work}

This paper presents an hourly Linear Quadratic MPC control strategy that can be effectively used to tackle the storage control problem for the Lake Como. LQ MPC is shown to be computationally efficient, satisfactory with respect to the main control objectives and comparable with offline non causal strategies such as DDP. The idea of formulating the dry avoidance objective as a hard constraint allowed us to decrease the complexity of the MPC tuning. A daily approximation of the hourly LQ MPC is developed to provide a satisfactory and comparable control system for cases in which only daily inflow data are available. 

Future work will be dedicated to the study of the stochastic MPC setting, where the impact of the uncertainty in the water inflow forecasts on the closed-loop performance can be properly assessed. 

\begin{comment}
\begin{table}[H]
\begin{center}
\caption{Daily vs Hourly MPC Demand Performances}\label{tb:Demand}
\begin{tabular}{|p{3.5cm}|p{2cm}|p{2cm}|}
\hline
&\textit{Daily} &\textit{Hourly} \\ \hline
\textit{RMSE Demand} $[m^3/s]$ & 40.876 & 40.342 \\
\hline
\textit{Deficit Peak} $[m^3/s]$& -118.390 &-116.480\\ 
\hline
\textit{Deficit Hours} & 2688 & 2448 \\
\hline
\textit{Area Deficit} $[m^3/s*hours]$ & 79244 & 71688 \\
\hline 
\end{tabular}
\end{center}
\end{table}
\end{comment}
\bibliography{ifacconf}             % bib file to produce the bibliography
                                                     % with bibtex (preferred)
                                                   
%\begin{thebibliography}{xx}  % you can also add the bibliography by hand

%\bibitem[Able(1956)]{Abl:56}
%B.C. Able.
%\newblock Nucleic acid content of microscope.
%\newblock \emph{Nature}, 135:\penalty0 7--9, 1956.

%\bibitem[Able et~al.(1954)Able, Tagg, and Rush]{AbTaRu:54}
%B.C. Able, R.A. Tagg, and M.~Rush.
%\newblock Enzyme-catalyzed cellular transanimations.
%\newblock In A.F. Round, editor, \emph{Advances in Enzymology}, volume~2, pages
%  125--247. Academic Press, New York, 3rd edition, 1954.

%\bibitem[Keohane(1958)]{Keo:58}
%\newblock \emph{Power and Interdependence: World Politics in Transitions}.
%\newblock Little, Brown \& Co., Boston, 1958.

%\bibitem[Powers(1985)]{Pow:85}
%\newblock Is there a way out?
%\newblock \emph{Harpers}, pages 35--47, June 1985.

%\bibitem[Soukhanov(1992)]{Heritage:92}
%\newblock \emph{{The American Heritage. Dictionary of the American Language}}.
%\newblock Houghton Mifflin Company, 1992.

%\end{thebibliography}
\begin{comment}
\appendix
\section{A summary of Latin grammar}    % Each appendix must have a short title.
\section{Some Latin vocabulary}              % Sections and subsections are supported  

\end{comment}                                                                     % in the appendices.
\end{document}